# Phase States and Phase Portraits of Tunnel Traffic. Empirical Data Analysis

## Analysis of Empirical Data Collected in Lefortovo Tunnel in Moscow

**Ihor Lubashevsky[1], Namik Guseyn-zade[1], Cyril Garnisov[2], Boris Livshits[2]**

[1]A.M. Prokhorov General Physics Institute, Russian Academy of Sciences, Moscow

[2]Moscow Technical University for Radioengineering, Electronics, and Automation, Moscow

**Abstract** The 3D fundamental diagrams and phase portraits for tunnel traffic is constructed based on the empirical data collected during the last years in the deep long branch of the Lefortovo tunnel located on the 3$^{rd}$ circular highway in Moscow. This tunnel of length 3 km is equipped with a dense system of stationary radiodetetors distributed uniformly along it chequerwise at spacing of 60 m. The data were averaged over 30 s. Each detector measures three characteristics of the vehicle ensemble; the flow rate, the car velocity, and the occupancy for three lanes individually. The conducted analysis reveals complexity of phase states of tunnel traffic. In particular, we show the presence of cooperative traffic dynamics in this tunnel and the variety of phase states different in properties. Besides, the regions of regular and stochastic dynamics are found and the presence of dynamical traps is demonstrated.

## Introduction

Traffic flow dynamics in long highway tunnels has been studied individually since the middle of the last century (see, e.g., (Chin and May 1991) or (Rothery 2001)). Interest to this problem is due to several reasons. The first and, may be, main one is safety. Jam formation in long tunnels is rather dangerous and detecting the critical states of vehicle flow leading to jam is of the prime importance for the tunnel operation. However, the tunnel traffic in its own right is also an attractive object for studying the basic properties of vehicle ensembles on highways. Indeed, on one hand, the individual car motion is more controllable inside tunnels with respect to velocity limits and lane changing. On the other hand, long tunnels typically are equipped well for monitoring the car motion practically continuously along them, which provides a unique opportunity to receive detailed information about the spatial-temporal structures of traffic flow.

In this paper we analyze the basic properties exhibited by tunnel congested traffic. The analysis is based on empirical data collected during the last years in the Lefortovo tunnel (Fig. 1) located on the 3-rd circular highway in Moscow. It comprises two branches and the upper one is a linear three lane tunnel of length about 3 km. Exactly in this branch the presented data were collected. The tunnel is equipped with a dense system of stationary radiodetetors distributed uniformly along it chequerwise at spacing of 60 m. Because of the detector technical features traffic flow on the left and right lanes is measured at spacing of 120 m whereas on the middle lane the special resolution gets 60 m. The data were averaged over 30 s.

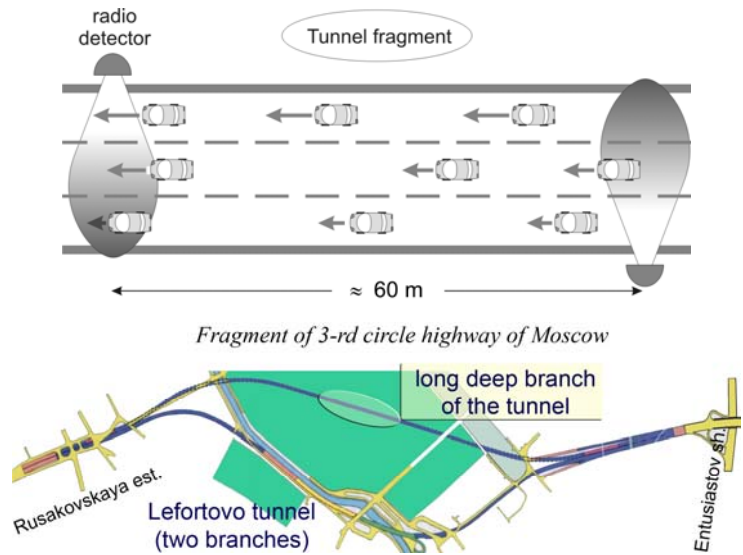

**Fig. 1.** The Lefortovo tunnel structure and the arrangement of radiodetectors in this tunnel.

Each detector measures three characteristics of the vehicle ensembles; the flow rate $q$, the car velocity $v$, and the occupancy $k$ for three lanes individually. The occupancy is an analogy to the vehicle density and is defined as the total relative time during with vehicles were visible in the view region of a given detector within the averaging interval. The occupancy is measured in percent.

## Cooperative motion of vehicle ensembles

First of all, this section demonstrates that the observed traffic flow indeed exhibits cooperative dynamics when the vehicle density becomes high enough. To do this figure 2 exhibits the spatial autocorrelations in the occupancy, car velocity, and flow rate measured by different detectors at the middle lane on 28.09.2005 when congested traffic was dominant. In agreement with the single-vehicle data (Neubert et al, 1991) the congested vehicle motion is characterized by essential correlations especially in the car velocity. The flow rate measurements are correlated substantially only within several neighboring detectors (on scales about several hundred meters) whereas the velocity measurements as well as the occupancy ones are correlated at half of the tunnel length, i.e. at scales about one kilometer.

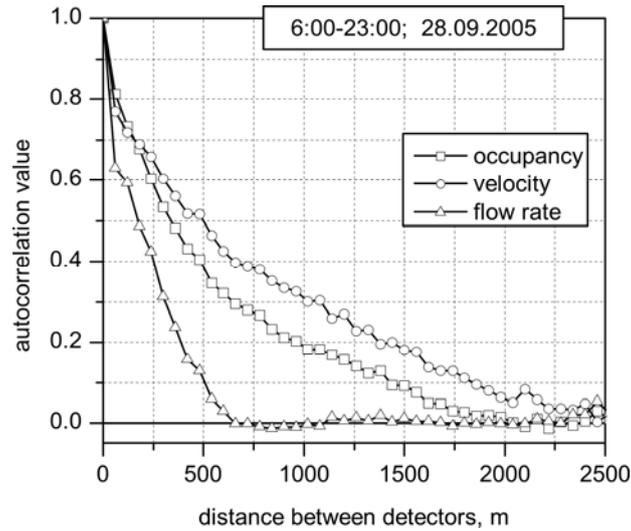

**Fig. 2.** Autocorrelations in the occupancy, car velocity, and flow rate measured by differing detectors vs the distance between them.

## 3D Fundamental Diagram

To analyze the phase states of tunnel traffic its fundamental diagram is studied in detail. The fundamental diagram under consideration was constructed as follows. The phase space $\{k,v,q\}$ was divided into cells of size about
$$1\% \times 1 \text{ km/h} \times 0.01 \text{ car/s}$$
Each 30 seconds a detector contributes unity to one of the cells. Taking into account a certain rather long time interval of traffic flow observation, all the detectors, and then dividing the result by the total number of records we obtain the three-dimensional distribution $P(k,v,q)$ of fixed traffic flow states over this phase space. In order to elucidate the obtained result we present the projection of $P(k,v,q)$ on three phase planes $\{kq\}$, $\{kv\}$, and $\{vq\}$. Besides, in projecting onto the given phase planes some layers can be singled out, for example, the expression

$$P_{DV}(k,q) \propto \int_{v \in DV} dv\, P(k,v,q)$$

specifies the projection of the layer $DV = (v_{min}, v_{max})$ onto the plane $\{kq\}$ within a constant cofactor normalizing it to unity. Such distributions will be also referred to as slices of the fundamental diagram.

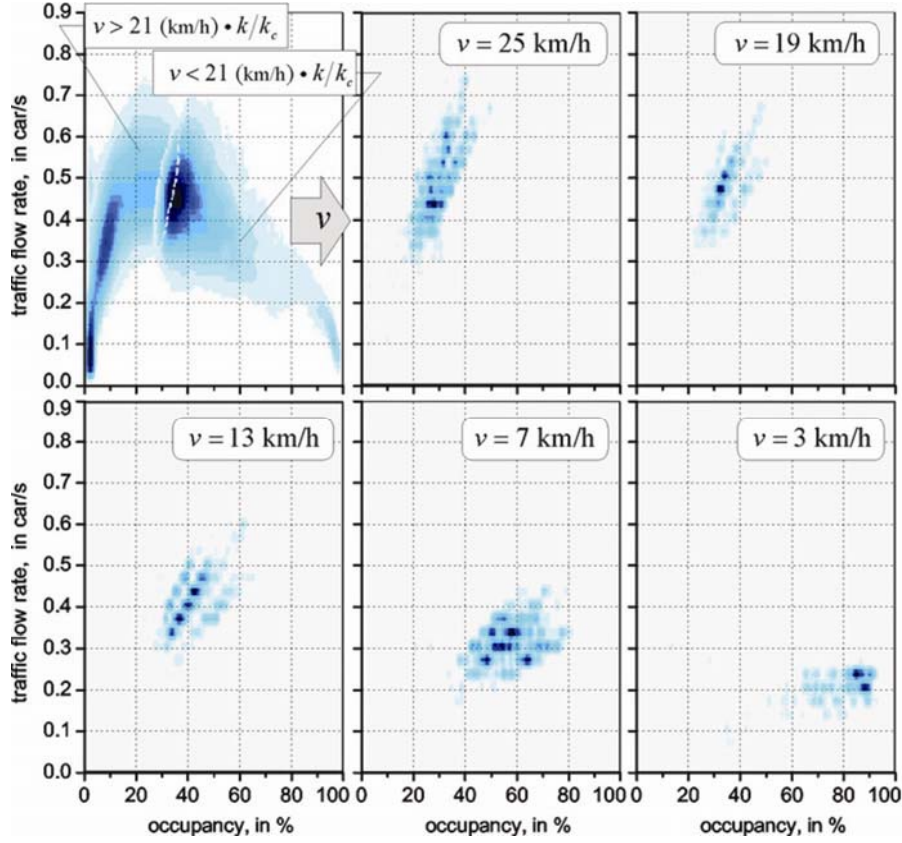

**Fig. 3.** Projection of the fundamental diagram onto the plane $\{k,q\}$ as well as its slices parallel to this plane.

Figure 3 presents the projection of the whole fundamental diagram onto the plane $\{k,q\}$ (the upper left frame) as well as its slices parallel to this plane. In the frame of the whole projection two branches are singled out by the relation $v \lesssim 21 \text{km/h} \times k/k_{c2}$, where the critical occupancy $k_{c2} = 31\%$ according the results to be demonstrated further. The two branches with a small degree of overlap are separated actually by the transition from light to heavy synchronized traffic (see below). The given slices of fixed velocity demonstrate the fact that, at least, three different states of heavy congested traffic were observed. It reflects in the existence of three branches visible well for *v* = 19, 13, 7 km/h. Their additional analysis demonstrated us that these branches are characterized individually by different mean lengths of vehicles. In particular, the higher is a branch in Fig. 3, the shorter, on the average, vehicles forming it. The distribution of the traffic flow states becomes rather uniform for very low velocities matching the jam formation. On the whole fundamental diagram the jammed traffic is described by the region looking like a certain "beak".

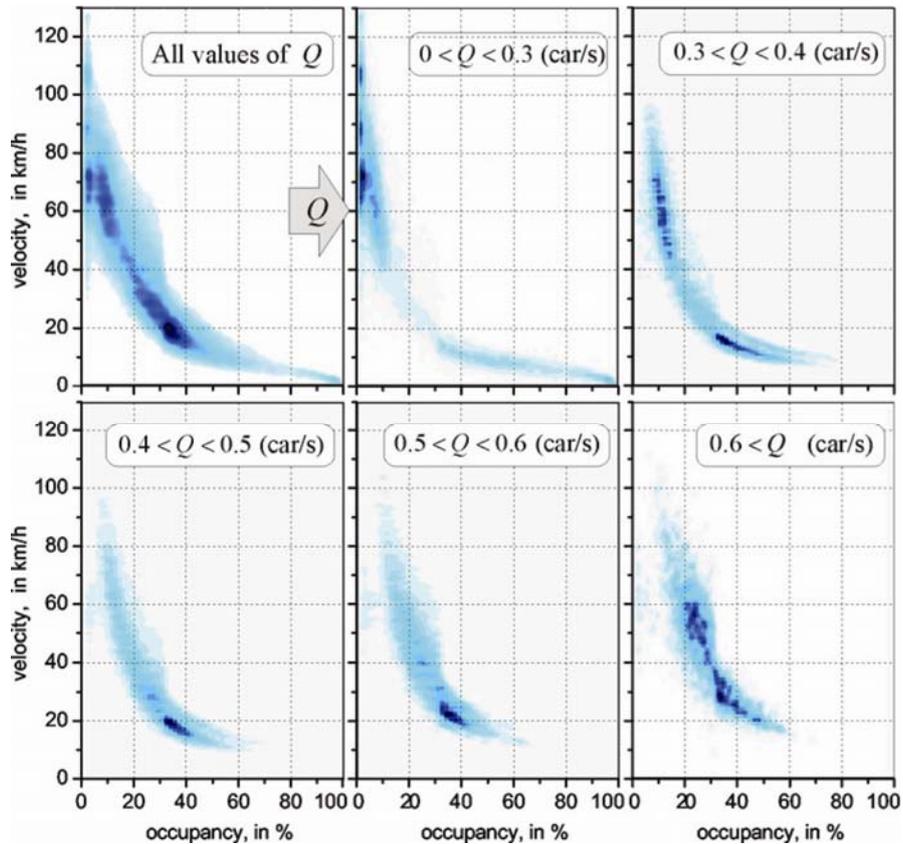

**Fig. 4.** Projection of the fundamental diagram onto the plane $\{k,v\}$ as well as its slices parallel to this plane that are made up by projecting the noted layers.

Figure 4 depicts a similar projection of the fundamental diagram onto the plane $\{k,v\}$. For low values of the traffic flow rate two states of traffic flow are clearly visible, the free flow and jam. The slice of $0.3 < q < 0.4$ (car/s) shows actually the light and heavy phases of synchronized traffic flow, with the latter phase splitting into several branches. The final slice corresponding to large values of the traffic flow rate exhibits the phase transition between the two light and heavy states of traffic flow at the critical value of occupancy $k_{c2} = 31\%$, where the velocity drop about 15 km/h is clearly visible. It should be pointed out that the traffic flow states are distributed with the comparable intensity on both the sides of the phase transition at $k = k_{c2}$, which enables us to assume that this phase transition proceeds in the both directions.

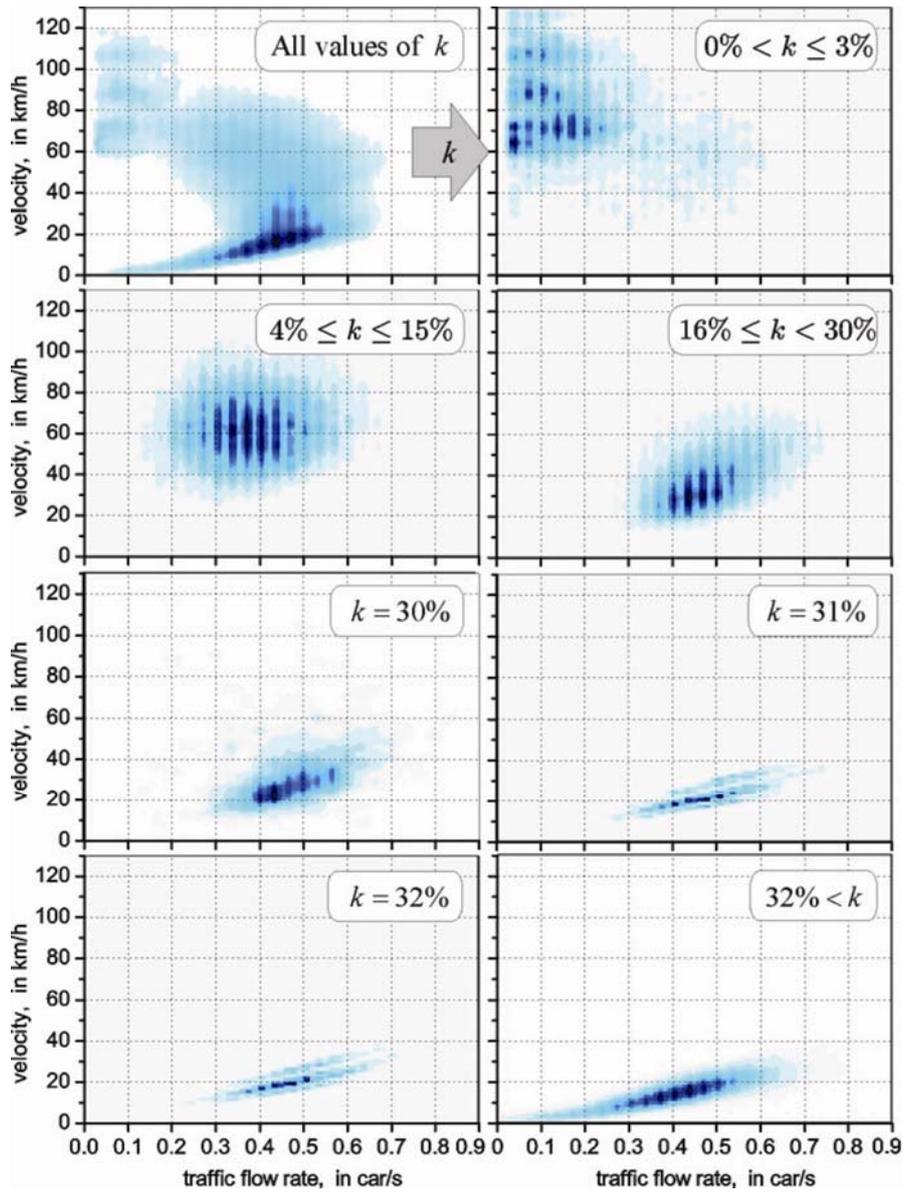

**Fig. 5.** Projection of the fundamental diagram onto the $\{q,v\}$-plane as well as its several slices parallel to this plane.

The whole projection of the fundamental diagram on the plane $\{k,v\}$ also shows this phase transition as well as the existence of two accumulation points of traffic flow states in the region of light synchronized traffic and in the vicinity of phase transition between the two states of synchronized traffic. The latter feature poses a question about the possibility of phenomena like "stop-and-go waves" but based on transitions between different states of the synchronized traffic.

Figure 5 exhibits the projection of the fundamental diagram onto the plane $\{q,v\}$ and evolution of its slices for fixed values of the occupancy. In this figure the four different phase states of the analyzed tunnel traffic are visible. The free flow where the overtaking maneuvers are most feasible corresponds to the three branches that can be related to trucks, passenger cars, and high-speed cars. As the traffic flow rate grows with the occupancy $k$ the three branches terminate and are followed by a structureless two-dimensional domain via a certain phase transition. Then this phase state in turn is followed by a structural domain which itself converts again into the structureless beaked region corresponding to jam.

The mechanism of the found substructures of the heavy synchronized traffic requires special investigations. Now we see two alternatives. One is the real existence of several individual states of the heavy tunnel traffic. This point of view is justified by the fact that the three branches were fixed for different lanes, for example, the low speed branch was observed only at the right and middle lanes. When vehicle density becomes sufficiently high the lane change maneuvers have to be depressed and, at the first approximation, traffic flow on different lanes is mutually independent. The other is that the observed splitting of the fundamental diagram is an artifact caused by the technical features of the detectors. The measured velocity can depend on the vehicle length. So in the reality we see only one branch. In any case the collected data set enables us to draw a conclusion that the lightly synchronized traffic is characterized by widely scatted states and for it there is no a direct relationship between the phase parameters $k$, $v$, $q$. Contrary, for the heavy synchronized traffic there is a relationship between them, for example,

$$q = Q(k,v).$$

It should be noted that only the consideration of 3D fundamental diagram exhibits this feature. In the projection, for example, on the $\{k,q\}$ the given surface is mapped on a 2D region.

## Portraits of tunnel vehicle dynamics in the phase space

The characteristics of the vehicle ensemble dynamics in the phase space $\{k,v\}$ were studied in the following way, replicating actually the technique by Friedrich et al 2002 used in a similar analysis. The plane $\{k,v\}$ is divided into cells $\{C\}$ of size $2.5\% \times 2.5\,\text{km/h}$. Let at time $t$ the traffic flow measurements of a given detector fall in a cell $C_i$ and in the averaging time $dt = 30\,\text{s}$ the next measurements of the same detector are located in a cell $C_j$. Then the vector $d\mathbf{r} := \{dk_t, dv_t\}$ such that $dk_t = k_j - k_i$ and $dv_t = v_j - v_i$ describes the system motion on the phase plane at the given point $\mathbf{r}_i := \{k_i, v_i\}$ at time $t$. These vectors were calculated using the data collected on 28.09.2005 by all the detectors. Averaging the found vectors gives the drift field $\mathbf{V}_m(\mathbf{r}) = \langle d\mathbf{r} \rangle / dt$ and the intensity $D(\mathbf{r})$ of an effective random force determined as

$$Ddt = \sqrt{\langle |d\mathbf{r}|\rangle^2 - \langle d\mathbf{r}\rangle^2}$$

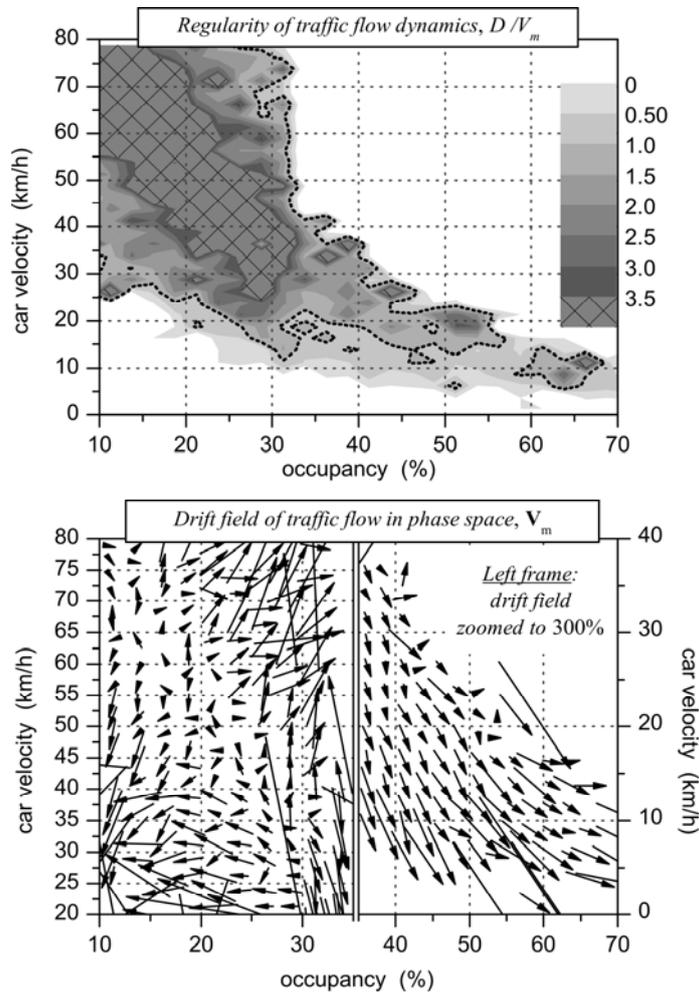

**Fig. 6.** Structure of ensemble vehicle dynamics in the phase space $\{k,v\}$. The upper window visualizes distribution of ratio between the random and regular components of the effective forces. The lower window depicts the regular drift field.

Figure 6 exhibits these fields. The upper window depicts the ratio $\eta := D/|\mathbf{V}|_m$, namely, its variations from 0 up to 3.5. The white region comprises the cells where no measurements were obtained. The hatched domain matches the ratio $\eta > 3.5$, where the vehicle ensemble dynamics can be regarded as pure random. The region between them contains several levels of the ratio $\eta$ variations and the level $\eta = 1.0$ is singled out in Fig. 6. For smaller values of $\eta$ the dynamics of vehicle ensemble becomes practically regular. The lower window of Fig. 6 shows the drift field $\mathbf{V}_m(\mathbf{r})$. Since its intensity changes essentially at different parts of the plane $\{k,v\}$ two frames are used to visualize it. In the left frame the drift field is zoomed in by three times relative to the right one. Let us consider them individually. The system dynamics in the right frame is rather regular and the filed $\mathbf{V}_m(\mathbf{r})$ corresponds to the irrelievable drift of vehicle ensemble to smaller velocities and higher densities. In other words, it is some visualization of the jam formation. In fact one or two jams were the case on that day. It should be noted that the transition region separating the left frame pattern being rather chaotic and the given one is relatively thin, it is located at $k = 35\%$ and has a thickness less than 5%,. So the observed jam formation seems to proceed via some breakdown in the cooperative vehicle motion, which is an agreement with other data (Kerner B.S. 2004).

The pattern shown in the left frame matches the upper one in structure. Inside a neighborhood $\mathcal{Q}_0$ of the decreasing frame diagonal the traffic dynamics is practically pure chaotic, at least, the found values of $\mathbf{V}_m(\mathbf{r})$ are relatively small and their directions do not form any regular pattern. As it must, outside this domain the field $\mathbf{V}_m(\mathbf{r})$ becomes more regular and the obtained data enable us to estimate its characteristic direction. The found structure of the drift velocity is rather anomalous. It looks like an one dimensional region (the diagonal $\mathcal{Q}_0$) of zero values of the drift velocity $\mathbf{V}_m(\mathbf{r})$ surrounded by regular system motion along it. The latter is clearly seen in the left low corner in Fig. 6 (left frame). Such behavior of a dynamical system can be explained using the notion of dynamical traps predicting also the existence of a long-lived state multitude as a consequence of some nonequilibrium phase transitions caused by the human bounded rationality (Lubashevsky et al, 2002, 2003, 2005).

Angle distribution of the vectors $\{d\mathbf{r}\}_i$ characterizing the system displacements on the analyzed phase plane at a given point $\{k,v\}_i$ typically contains two local maxima. Mutually they specify the field $\mathbf{V}_m(\mathbf{r})$. Keeping in mind the fact that the 3D phase space $\{k,v,q\}$ gives a more adequate description of the vehicle dynamics the pattern of these maxima was studied individually. Figure 7 visualizes the obtained results. Roughly speaking, the two patterns describe the regular system motion in the opposite directions on the phase plane $\{k,v\}$. Again the region with $k > 30\% - 33\%$ exhibits rather strong regular motion towards high dense flow.

However, the regular motion in the opposite direction also can be intensive for $k < 30\%$. It is natural to relate the two fields to the hysteresis effect in the jam formation observed in traffic. The found patterns seem to be formed by different slices of the whole phase space $\{k,v,q\}$, which is worthy of individual consideration.

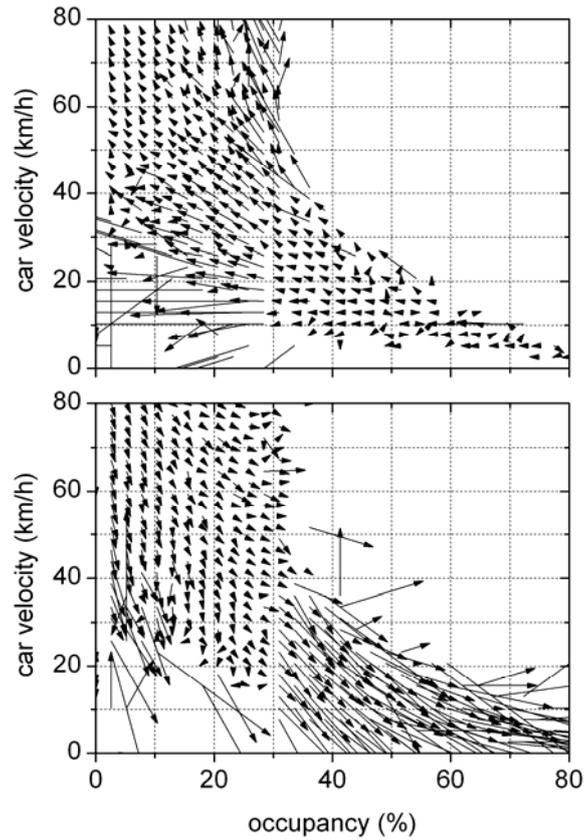

**Fig. 7.** Two components of the system regular motion on the plane $\{kv\}$ describing, roughly speaking, the regular system motion in the opposite directions.

## Conclusion

The present work is devoted to constructing the 3D fundamental diagram for tunnel traffic and its phase portraits based on the empirical data collected in the linear higher branch of the Lefortovo tunnel located on the 3$^{rd}$ circular highway in Moscow during 2004-2005. It is the three lane tunnel of length 3 km equipped with radiodetectors measuring the traffic flow rate ($q$, in car/s), the vehicle velocity ($v$, in km/h), and the road occupancy ($k$, in \%) averaged over 30 s. The detectors are distributed chequerwise at spacing 60 m along the tunnel. Because of the detector technical characteristics the traffic flow parameters are fixed at 60 m spacing on the middle lane and 120 m spacing on the left and right ones.

First of all, it is shown that the observed tunnel congested traffic in fact exhibits cooperative phenomena in vehicle motion. Indeed, the spatial autocorrelations in the occupancy, vehicle velocity, and flow rate measured by different detectors are found to be essential. Especially it concerns the correlations in the velocity and occupancy, their correlation length gets values about 1 km. The occupancy data are correlated on substantially shorter scales about 200-300 m.

The fundamental diagram is treated as the traffic flow state distribution and has been constructed using the relative number of records per
$$1\% \times 1 \text{ km/h} \times 0.01 \text{ car/s}$$
cells in the phase space $\{k,v,q\}$. Analyzing the three projections of this 3D field and its different slices we have demonstrated the fundamental diagram to be complex in structure. Four possible traffic flow states are found, the free flow, light synchronized traffic, heavy synchronized traffic, and jam. The free flow state as well as the heavy synchronized traffic has a substructure, whereas the light synchronized traffic and jam are structerless.

Based on the obtained results it seems that the free flow comprises three branches related to trucks, passenger cars, and high-speed cars. These branches exist while the occupancy is less than a certain critical value, $k < k_{c1} \approx 3\%$, and are clearly visible in the projection onto the phase plane $\{q,v\}$. As the occupancy grows the light synchronized traffic appears which is characterized by the structureless region of widely scatted states. When the occupancy exceeds the next critical value $k_{c2} \approx 31\%$ the heavy synchronized traffic changes the previous phase state. This transition is accompanied by some jump in the mean velocity.

In the projections onto the phase planes $\{k,q\}$, $\{k,v\}$, and $\{q,v\}$ it looks like widely scatted states uniformly distributed inside a certain region. However the corresponding slices of the fundamental diagram demonstrate a substructure of the given phase state. It again comprises, at least, three different branches. The conducted analysis demonstrated that the given branches are characterized, on the averaged, by different lengths of vehicles. Roughly speaking, the heavy synchronized traffic is characterized by a certain relationship between the occupancy, the mean velocity of vehicles, and traffic flow rate, for example, $q = q(k,v)$.

The jam phase, as should be expected, can be ascribed with a certain relationship between the traffic flow rate $q$, the mean velocity $v$, and the occupancy $k$, individually, in particular, it is possible to write down a certain function $v = v(k)$ and, thus, $q(k) = kv(k)$.

In addition we should note the following. In spite of the complex structure of the fundamental diagram and the existence of four different phase states the distribution of the detected states is, roughly speaking, bimodal. One its maximum is located at the beginning of the region matching the light synchronized traffic. The other maximum drops on the region corresponding to the transition between the two phases of the synchronized traffic.

The phase portrait of the vehicle ensemble dynamics on the occupancy-velocity plane is studied. It is demonstrated that there are two substantially different regions on it. One matches actually the cooperative vehicle motion and contains some kernel where the dynamics is pure chaotic. The other part of the phase plane corresponds to the irreversible stage of jam formation. The two regions are separated by a rather narrow transition layer located at k ~ 30%, which demonstrates that the observed jams originated inside a congested traffic via some breakdown. The constructed bimodal pattern of the system regular dynamics on the phase plane $\{k,v\}$ matches the hysteresis effects in the jam formation.

**Acknowledgments** This work was supported in part by DFG project MA 1508/8-1 and RFBR grant 06-01-04005.

**References**

Chin H. C. & May A. D. (1991). Examination of the speed-flow relationship at the Caldecott tunnel. In: *Transportation Research Record* **1320**, (pp.1-15), Transportation Research Board, NRC, Washington, DC.

Friedrich R., Kriso S., Peinke J., Wagner P. (2002), *Phys. Lett. A* **299**, 287.

Lubashevsky I., Mahnke R., Wagner P., and Kalenkov S. (2002). *Phys. Rev. E* **66**, 016117.

Lubashevsky I., Hajimahmoodzadeh M., Katsnelson A., Wagner P. (2003). *Eur. Phys. J. B* **36**, 115.

Lubashevsky I., Mahnke R., Hajimahmoodzadeh M., Katsnelson A. (2005). *Eur. Phys. J. B* **44**, 63.Kerner B.S. (2004), *Physics of traffic flow*, Springer, Berlin.

Neubert L., Santen L., Schadschneider A., Schreckenberg M. (1999), *Phys. Rev. E* **60**, 6480.

Rothery R. W. (2001). Car following models. In: N. Gartner, C. J. Messer, A. K. Rathi (Ed.) *Traffic Flow Theory*, Transportation Research Board, Special Report **165**, Chap. 4.